\documentclass[prb,showpacs,preprintnumbers,amsfonts,amssymb,amsmath,floats,twocolumn,aps]{revtex4}

\usepackage{graphicx}
\usepackage{dcolumn}
\usepackage{bm}
\usepackage[final,dvips]{epsfig}
\usepackage{bm}
\usepackage{color}
\usepackage{soul}

\DeclareMathAlphabet\mathbfcal{OMS}{cmsy}{b}{n}
\newcommand{\ff}[1]{{\boldsymbol #1}}
\newcommand{\ca}[1]{{\cal #1}}

\begin{document} 
   
\title{Strong-coupling limit of depleted Kondo- and Anderson-lattice models}

\author{Irakli Titvinidze, Andrej Schwabe and Michael Potthoff}

\affiliation{I. Institut f\"ur Theoretische Physik, Universit\"at Hamburg, Jungiusstra\ss{}e 9, 20355 Hamburg, Germany}

\begin{abstract}
Fourth-order strong-coupling degenerate perturbation theory is used to derive an effective low-energy Hamiltonian for the Kondo-lattice model with a depleted system of localized spins.
In the strong-$J$ limit, completely local Kondo singlets are formed at the spinful sites which bind a fraction of conduction electrons. 
The low-energy theory describes the scattering of the excess conduction electrons at the Kondo singlets as well as their effective interactions generated by virtual excitations of the singlets.
Besides the Hubbard term, already discussed by Nozi\`eres, we find a ferromagnetic Heisenberg interaction, an antiferromagnetic isospin interaction, a correlated hopping and, in more than one dimensions, three- and four-site interactions.
The interaction term can be cast into highly symmetric and formally simple spin-only form using the spin of the bonding orbital symmetrically centered around the Kondo singlet.
This spin is non-local. 
We show that, depending on the geometry of the depleted lattice, spatial overlap of the non-local spins around different Kondo singlets may cause ferromagnetic order. 
This is sustained by a rigorous argument, applicable to the half-filled model, by a variational analysis of the stability of the fully polarized Fermi sea of excess conduction electrons as well as by exact diagonalization of the effective model.
A similar fourth-order perturbative analysis is performed for the depleted Anderson lattice in the limit of strong hybridization. 
Even in a parameter regime where the Schrieffer-Wolff transformation does not apply, this yields the same effective theory albeit with a different coupling constant.
\end{abstract} 
 


\maketitle 

\section{Introduction}

The Kondo-lattice model \cite{Don77,LC79,TSU97b,KK00,Col07} is a prototypical model of itinerant conduction electrons interacting with a system of localized magnetic moments. 
It is a generic model to describe, e.g., the magnetism of heavy-fermion systems where the localized magnetic moments result from a partially filled inner shell and where those moments are coupled indirectly by means of the conduction electrons. 
The indirect coupling is caused by a local, typically antiferromagnetic exchange interaction of the form $J \ff s_{i} \ff S_{i}$ where $\ff S_{i}$ is the localized spin at a site $i$ and where $\ff s_{i}$ is the local spin of the conduction-electron system at the same site.
If $J$ is sufficiently weak, the effective interaction $J_{\rm RKKY} \propto J^{2}$ between the localized spins can be derived perturbatively. 
\cite{RK54,Kas56,Yos57}

For moderate $J$, collective magnetism competes against the Kondo effect, \cite{Wil75,Hew93} i.e., the screening of a localized spin by an extended cloud of conduction electrons, and in the absence of collective magnetic order, the scattering of the conduction electrons at the localized spins leads to the formation of a strongly correlated heavy-fermion state. 
But even in the strong-coupling regime, an indirect magnetic coupling between the localized spins survives.
For example, an extended range of ferromagnetic order in the phase diagram has been found for the $D=1$ dimensional model \cite{TW93,TSU97b,McCJRG02,Gul04,PK12} or for the model on $D=\infty$ dimensional lattices. \cite{OKK09,PKP12}
Here, the Kondo effect has been recognized to even cooperate with magnetic ordering.
\cite{PK12,PKP12}
Furthermore, for $J\to\infty$ the problem can be mapped onto the infinite-$U$ Hubbard model by identifying unscreened spins with singly occupied sites and local Kondo singlets with unoccupied sites. \cite{Lac85}
Therewith, ferromagnetism in the strong-$J$ and low-electron-density limit can be related to the Nagaoka mechanism. \cite{Nag66}

The essence of the Kondo effect is actually captured by the Kondo-impurity model. \cite{Kon64,Hew93} 
The impurity case can be realized by a single ($R=1$) localized spin which is antiferromagnetically exchange coupled to a system of $N$ conduction electrons hopping over a lattice of $L$ sites.
Contrary, there are $R=L$ localized spins in the Kondo-lattice model.
Here the following question suggests itself and is obviously of great fundamental importance:
How does ferromagnetic order emerge on the way from the impurity case $R=1$ (non-magnetic), over the dilute case with a small fraction $R/L$ of magnetic impurities, to the dense case with $R=L$?

An important model in this context is the depleted Kondo-lattice model with a number of $R<L$ localized spins which is still far from the dilute limit.
For a regular depletion of the lattice of localized spins, with a certain fixed spin-spin distance $d$, this model fully comprises the intricate physics of local or temporal quantum fluctuations present in the impurity case.
However, the full complexity of lattice coherence effects is somewhat suppressed and, depending on $d$, the model is more accessible to a mean-field-like picture with less important spatial correlations. \cite{STP13}

\begin{figure}[t]
\centerline{\includegraphics[width=0.37\textwidth]{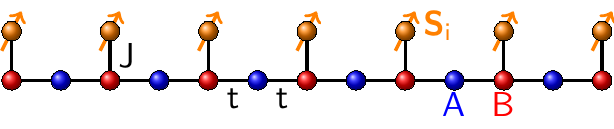}  }
\vspace{3mm}
\centerline{\includegraphics[width=0.37\textwidth]{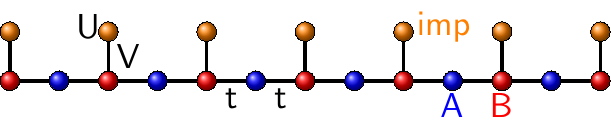}  }
\caption{(Color online)
Schematic picture of a depleted Kondo-lattice (top) and a depleted Anderson-lattice model (bottom) in $D=1$ dimension with a distance $d=2$ between localized spins or orbitals, respectively.  
Nearest-neighbor hopping $t$ of conduction electrons takes place between the sites of two sublattices $A$ and $B$ (blue and red). 
In the Kondo case, the $B$ sublattice sites are coupled via an antiferromagnetic local spin exchange $J$ to the localized spins (yellow). 
In the Anderson case, the $B$ sublattice sites are coupled by a hybridization $V$ to the correlated impurity sites (yellow) with local Hubbard interaction $U$.
The conduction-electron system consists of $L$ lattice sites and $N$ electrons. 
There are $R$ local spins or localized orbitals, respectively. 
}
\label{fig:depleted_1D}
\end{figure}

It still mimics a metallic heavy-fermion state:
Starting from a Kondo insulator that can be realized on a dense and half-filled Kondo lattice, $N=R=L$, a metallic heavy-fermion state 
is usually approached by ``doping'' the system, $N<L$.
Depletion of the lattice of localized spins, $R<L$, can likewise lead to a metallic state, as pointed out in Ref.\ \onlinecite{Ass02}, as this produces excess electrons which, in the strong $J$ limit, are not bound in local Kondo singlets but are itinerant.
For the $D=1$ dimensional case, a sketch of the depleted Kondo lattice model with a distance $d=2$ between the localized spins, i.e., $R=L/2$ is given by Fig.\ \ref{fig:depleted_1D} (top).

With the present study we address the depleted Kondo-lattice model in the strong-coupling limit. 
Our main goal is to derive an effective low-energy Hamiltonian by means of strong-coupling perturbation theory, i.e., perturbation theory in powers of $t/J$ where $t$ is the nearest-neighbor hopping connecting to the local Kondo singlets.
It turns out that fourth order perturbation theory is sufficient to lift the macroscopic ground-state degeneracy of the unperturbed $t=0$ Hamiltonian. 
The resulting effective Hamiltonian $\ca H_{\rm eff}$ describes the emergent correlations among the excess conduction electrons, which are {\em a priori} uncorrelated, resulting from virtual excitations of the local Kondo singlets as well as the scattering from the singlets.
$\ca H_{\rm eff}$ contains a Hubbard-like term as already predicted by Nozi\`eres \cite{Noz74,Noz76,NB80} but also includes additional non-local interaction terms.
Interestingly, it can be written in an extremely compact and formally simple form using a representation with non-local spins centered at the local Kondo singlets.

The benefit of the effective theory is that the strong-coupling physics of the depleted model can be addressed easily while direct numerical approaches, such as density-matrix renormalization, \cite{Whi92,Sch11} suffer from the necessity to resolve the extremely small energy gaps that become relevant in the strong-coupling limit.

We first consider the depleted Kondo lattice with spin-spin distance $d=2$ but then also for other distances $d\ge 2$ and for irregular depletion. 
Furthermore, the perturbation theory is carried out for an arbitrary $D$-dimensional lattice. 
Finally, we also consider the depleted Anderson lattice model in the limit of strong hybridization (see Fig.\ \ref{fig:depleted_1D}, bottom, for a sketch). 
This can be treated analogously -- also in a parameter regime where it cannot mapped onto the depleted Kondo lattice by means of the Schrieffer-Wolff transformation. \cite{SW66,SN02}

Fourth-order perturbation theory is sufficient if $d\ne 1$, i.e., if two localized spins (orbitals) are not nearest neighbors.
This somewhat restricts the conceivable geometries, particularly for dimensions $D>1$.
Our perturbative analysis applies to systems with conduction-electron concentrations $N/L$ such that there are excess conduction electrons that are not bound in local Kondo (Anderson) singlets for $J\to \infty$ ($V\to \infty$). 
This includes the case of a half-filled conduction band with $N/L=1$. 
In general, $N/L > R/L$ must be assumed. 
If $d=2$, for example, the case $N/L=R/L=1/2$ corresponds to a quarter-filled conduction band but the physics resembles the Kondo-insulator physics of the half-filled Kondo lattice in the dense case ($d=1$ or $N=R=L$). 
For concentrations $N/L < R/L$, Kondo singlets must be broken even in the unperturbed state.
The physics in the situation resembles the case of the doped Kondo insulator in the dense Kondo-lattice model.
Note that this regime has been analyzed by means of strong-coupling perturbation theory before, see Refs.\ \onlinecite{TSU97b,STUR92}.

The effective theory derived in the present paper has already been successfully used in different previous studies. 
Particularly, it has been employed to explain ferromagnetic correlations and ferromagnetic long-range order in different one- and two-dimensional realizations of the depleted Kondo model. \cite{STP13,TSP14} 
It explains the ferromagnetic coupling of local magnetic moments that are formed in the {\em a priori} uncorrelated conduction electron system on the $A$-sublattice sites (see Fig.\ \ref{fig:depleted_1D}) due to quantum confinement between local Kondo singlets, an effect that has been termed ``inverse indirect magnetic exchange''. \cite{STP13} 
Furthermore, the effective model has been employed in a recent work on exchange mechanisms in confined Kondo systems. \cite{SHP14}

There are many other possible fields of applications. 
For example, a closely related problem is that of superlattices consisting of a periodic arrangement of $f$-electron- and non-interacting two-dimensional layers studied in Ref.\ \onlinecite{PTK13}. 
Another type of systems is given by magnetic atoms on non-magnetic metallic surfaces where scanning-tunneling techniques nowadays not only allow for a manipulation the system geometry on the atomic scales but also an atomically precise mapping of spin-dependent couplings. \cite{ZWL+10,KWCW11,KWC+12}
It will also be interesting to employ the effective theory for studies of {\em randomly} depleted Kondo lattices which capture essential physical properties of Kondo alloys and have been studied previously, see Refs.\ \onlinecite{KV07,BL13}. 

In the present paper, the effective theory is used to study the magnetic properties of the model depicted by Fig.\ \ref{fig:depleted_1D}. 
For the case of half-filling and using the non-local spin representation, one can easily prove that the fully polarized ferromagnetic state is among the ground states.
Off half-filling we employ a simple variational wave-function approach as well as exact diagonalization to study the question whether the one-dimensional depleted Kondo lattice with $d=2$ sustains ferromagnetic order in the strong-$J$ regime. 

The paper is organized as follows:
The perturbation theory for the depleted Kondo lattice is worked out in the next section.
Sec.\ \ref{sec:and} deals with the more complicated depleted Anderson model.
A comparison between both is made in Sec.\ \ref{sec:comp}. 
In Secs.\ \ref{sec:spiniso} and \ref{sec:nonlocal} different representations of the effective model are discussed. 
The variational and exact-diagonalization results are presented in Sec.\ \ref{sec:analysis}.
Finally, Sec.\ \ref{sec:dilute} discusses the generalization to arbitrary, e.g., diluted system geometries. 
The main conclusions are summarized in Sec.\ \ref{sec:con}.

\section{Strong-coupling perturbation theory for the depleted Kondo-lattice model}
\label{sec:kondo}

The Hamiltonian of the depleted Kondo-lattice model is given by:
\begin{equation}
\label{Hamiltonian_Kondo} 
{\cal H} = t \sum_{\langle ij\rangle, \sigma} c_{i,\sigma}^\dagger c_{j,\sigma}^{\phantom\dagger}+ J \sum_{i \in B}{\bf s}_i{\bf S}_i \: .
\end{equation}
We consider a $D$-dimensional lattice consisting of $L$ sites labelled by $i$.
The first term describes the nearest-neighbor hopping of a system of $N$ non-interacting conduction electrons with hopping amplitude $t$ on this lattice.
$\sigma=\uparrow,\downarrow$ indicates the spin projection. 
The second term represents the local antiferromagnetic spin interaction with coupling strength $J>0$ at the sites of a sublattice $B$.

The sublattice $B$ consists of $R$ sites. 
The remaining $L-R$ sites of the lattice form the sublattice $A$. 
We do not assume the original lattice as bipartite, and the number of sites in $A$ and $B$ is arbitrary and may be different in particular.
It is required, however, that each $B$-sublattice site is connected by the nearest-neighbor hopping terms to $A$-sublattice sites only, i.e., sites with local interaction $J$ must be surrounded by uncorrelated sites.
The system given by Fig.\ \ref{fig:depleted_1D} (top) in fact represents the ``most dense'' system under this constraint.

The interaction term at a site $i \in B$ involves the local conduction-electron spin,
\begin{equation}
\label{spin}
{\bf s}_i 
=
\frac{1}{2} 
\sum_{\sigma\sigma'}
c_{i,\sigma}^\dagger
{\boldsymbol\sigma}_{\sigma\sigma'}
c_{i,\sigma'} \; ,
\end{equation}
where $\ff \sigma$ is the vector of Pauli matrices, and the localized spin $\ff S_{i}$. 
We assume that $\ff S_{i}$ is a spin-$1/2$ operator with $\ff S_{i}^{2} = 3/4$ as usual.

To derive an effective low-energy model for the strong-coupling limit (${|t| \ll J}$), we employ fourth-order perturbation theory in the hopping connecting the $B$-sublattice to the $A$-sublattice sites. 
For the system given by Fig.\ \ref{fig:depleted_1D} (top) this comprises {\em all} hopping terms, and we will first concentrate on this special situation for clarity.
The theory can easily be extended to more general geometries as well (see Sec.\ \ref{sec:dilute}).
 
As the starting point we first consider the unperturbed Hamiltonian 
\begin{equation}
\label{H0_Kondo}
{\cal H}_0= \sum_{i \in B}{\cal H}_0^{(i)}=J \sum_{i \in B} {\cal \bf s}_{i} {\cal \bf S}_i \, . 
\end{equation}
This is just the atomic limit ($t=0$) of the depleted Kondo lattice given by Fig.\ \ref{fig:depleted_1D} (top).
The antiferromagnetic interaction between impurity and conduction-electron spins on the $B$ sites favors the formation of local Kondo singlets:
\begin{equation}
\label{Kondo_singlet}
|{\rm KS}_i\rangle =\frac{1}{\sqrt{2}}\left(c_{i,\downarrow}^\dagger|0_{i}\rangle  \otimes |\uparrow _i\rangle -c_{i,\uparrow}^\dagger |0_{i}\rangle \otimes |\downarrow _i\rangle \right) \: .
\end{equation}
Here $|0_i\rangle$ denotes vacuum state at the $i$-th lattice site of the conduction-electron system, while $|M_i\rangle= {|\uparrow_i\rangle},{|\downarrow_i\rangle}$ refers to the spin state of the impurity that is coupled via $J$ to the $i$-th lattice site. 
For a total conduction-electron number $N$ in the range ${R < N < 2L-R}$ all impurity spins form Kondo singlets with conduction electrons.
The remaining $N-R$ electrons occupy $L-R$ sites. 
Each of the four possible ``atomic'' configurations 
${|X_j\rangle=|0_{j}\rangle, c_{j,\uparrow}^\dagger |0_{j}\rangle, c_{j,\downarrow}^\dagger |0_{j}\rangle, c_{j,\uparrow}^\dagger c_{j,\downarrow}^\dagger |0_{j}\rangle}$ can be realized at these sites 
since all sites are decoupled for $t=0$.
Consequently, there are $\frac{(2(L-R))!}{(N-R)!(2L-R-N)!}$ orthogonal ground states
\begin{equation}
\label{GS_H0}
|\phi_0\rangle=\bigotimes_{i \in B} |{\rm KS}_i\rangle \bigotimes_{j \in A} |X_j \rangle
\end{equation}
of the unperturbed Hamiltonian ${\cal H}_0$.

On the contrary, if the total electron number is $N<R$ or $N>2L-R$, not all impurity spins can form Kondo singlets with conduction-electron spins. 
There are either not enough or too many conduction electrons to screen all spins.
If $N<R$ the ground-state degeneracy is $\frac{R!}{N! (R-N)!}$ while for $N>2L-R$ the degeneracy is 
$\frac{R!}{(N-(2L-R))! (2L-N)!}$.

The ground-state degeneracy is expected to be lifted, partially or completely, apart from the trivial $2S_{\rm tot}+1$ spin degeneracy, by switching on the hopping $t$ between neighboring conduction-electron sites, i.e.\ by adding
\begin{equation}
\label{H1}
{\cal H}_1=t{\tilde{\cal H}}_1=t\sum_{i \in B} {\tilde{\cal H}}_1^{(i)}=t \sum_{i \in B} \sum_{j \in A_i} \sum_\sigma \left(c_{i,\sigma}^\dagger c_{j,\sigma} + \mbox{H.c.} \right)  \, .
\end{equation}
Here $A_i$ is the set of sites of sublattice $A$ neighboring the $B$-sublattice site $i$.

We aim at an effective low-energy Hamiltonian for the case ${R < N < 2L-R}$.
For the cases $N<R$ and $N>2L-R$ one needs to perform a different type of the perturbation theory. 
The latter would be similar to the one for the dense Kondo lattice carried out in Ref.\ \onlinecite{TSU97b}, and is not part of the present work.

To derive the effective Hamiltonian we employ standard perturbation theory as described in Ref.\ \onlinecite{EFG+05}, for example, in particular the following relation:
\begin{equation}
\label{perturbation_0}
P_0{\tilde{\cal H}}_1\sum_{k=0}^\infty t^{k+1} \left(\sum_{l>0}\frac{P_l {\tilde{\cal H}}_1}{E-E_l}\right)^k P_0  |\psi\rangle =(E-E_0)P_0|\psi\rangle \, .
\end{equation}
This is just a reformulation of the Schr\"odinger equation ${\cal H}|\psi\rangle=E|\psi\rangle$, with ${\cal H}={\cal H}_0 +t{\tilde{\cal H}}_1$, which is suitable for applying $k$-th order degenerate perturbation theory. 
In Eq.\ (\ref{perturbation_0})
\begin{equation}
\label{P0_Kondo}
P_0=\bigotimes_{i \in B} P_0^{(i)}  \quad {\rm with} \quad  P_0^{(i)}=|{\rm KS}_i\rangle \langle {\rm KS}_i| 
\end{equation}
denotes the projection operator onto the subspace of ground states of the unperturbed Hamiltonian with ground-state energy $E_0=-\frac{3}{4}JR$, while $P_l$ is the projection operator onto the $l$-th subspace of excited states with eigenenergy $E_l$. 
We have $P_l=\sum_\gamma |\phi_{l,\gamma}\rangle\langle \phi_{l,\gamma}|$ where the sum runs over all excited eigenstates
$|\phi_{l,\gamma}\rangle$ with the same energy $E_l$ which differ from each other by excitations of different local Kondo singlets. 

To derive the effective Hamiltonian, one has to get rid of the energy dependence of the left-hand side of Eq.\ \eqref{perturbation_0}. 
This is achieved by expanding the eigenenergy $E$ on the left-hand side in a power series, $E=E_0+\sum_{k=1}^\infty t^k E_0^{(k)}$.
Here $E_0^{(k)}$ are the $k$-th order corrections to the ground-state energy which can be determined order by order. 

The first excited energy level $E_1=E_0+3J/4$ corresponds to states where one of the local Kondo singlets is broken by adding or removing an electron from the corresponding $B$-sublattice site via virtual hopping processes.  
The according projection operator is
\begin{equation}
\label{P1}
P_1=\sum_{i \in B} P_1^{(i)} \bigotimes_{i' \in B}^{ i'\not= i} P_0^{(i')}  \; {\rm with}  \;
P_1^{(i)}  = |0_{i}\rangle\langle 0_i|  +|D_{i}\rangle\langle D_{i}|  \, .  
\end{equation}
Here $|D_{i}\rangle=c_{i,\uparrow}^\dagger c_{i,\downarrow}^\dagger |0_{i}\rangle$ stands for a doubly occupied site $i$.

States where one of the local Kondo singlets is excited to a local triplet state belong to the second excited energy level $E_2 = E_0 + J$.
The corresponding projection operator can be written as
\begin{equation}
\label{P2}
P_2=\sum_{i \in B} P_2^{(i)} \bigotimes_{i' \in B}^{i'\not=i} P_0^{(i')}  
\; {\rm with} \;  P_2^{(i)} = \sum_{m=-,0,+}  |T^m_i\rangle \langle T^m_i |  \, .
\end{equation}
Here $|T^{+}_i\rangle =c_{i,\uparrow}^\dagger|0_{i}\rangle  \otimes |\uparrow _i\rangle$, $|T^{-}_i\rangle =c_{i,\downarrow}^\dagger|0_{i}\rangle  \otimes |\downarrow _i\rangle $  and
$|T^0_i\rangle =\frac{1}{\sqrt{2}}\left(c_{i,\downarrow}^\dagger|0_{i}\rangle  \otimes |\uparrow _i\rangle + c_{i,\uparrow}^\dagger |0_{i}\rangle \otimes |\downarrow _i\rangle \right)$ are the triplet states. 

The third excited energy level contains two broken Kondo singlets, and its energy is $E_3=E_0+3J/2$. 
The corresponding projection operator is
\begin{equation}
\label{P3}
P_3=\sum_{i_1, i_2 \in B}^{i_1 > i_2}  P_1^{(i_1)} P_1^{(i_2)} \bigotimes_{i' \in B}^{i'\not=i_1, i_2} P_0^{(i')}  \, .
\end{equation}
This is the highest excited-state manifold that has to be taken into account in fourth-order perturbation theory.

\begin{figure}[bpt]
\centerline{\includegraphics[width=0.4\textwidth]{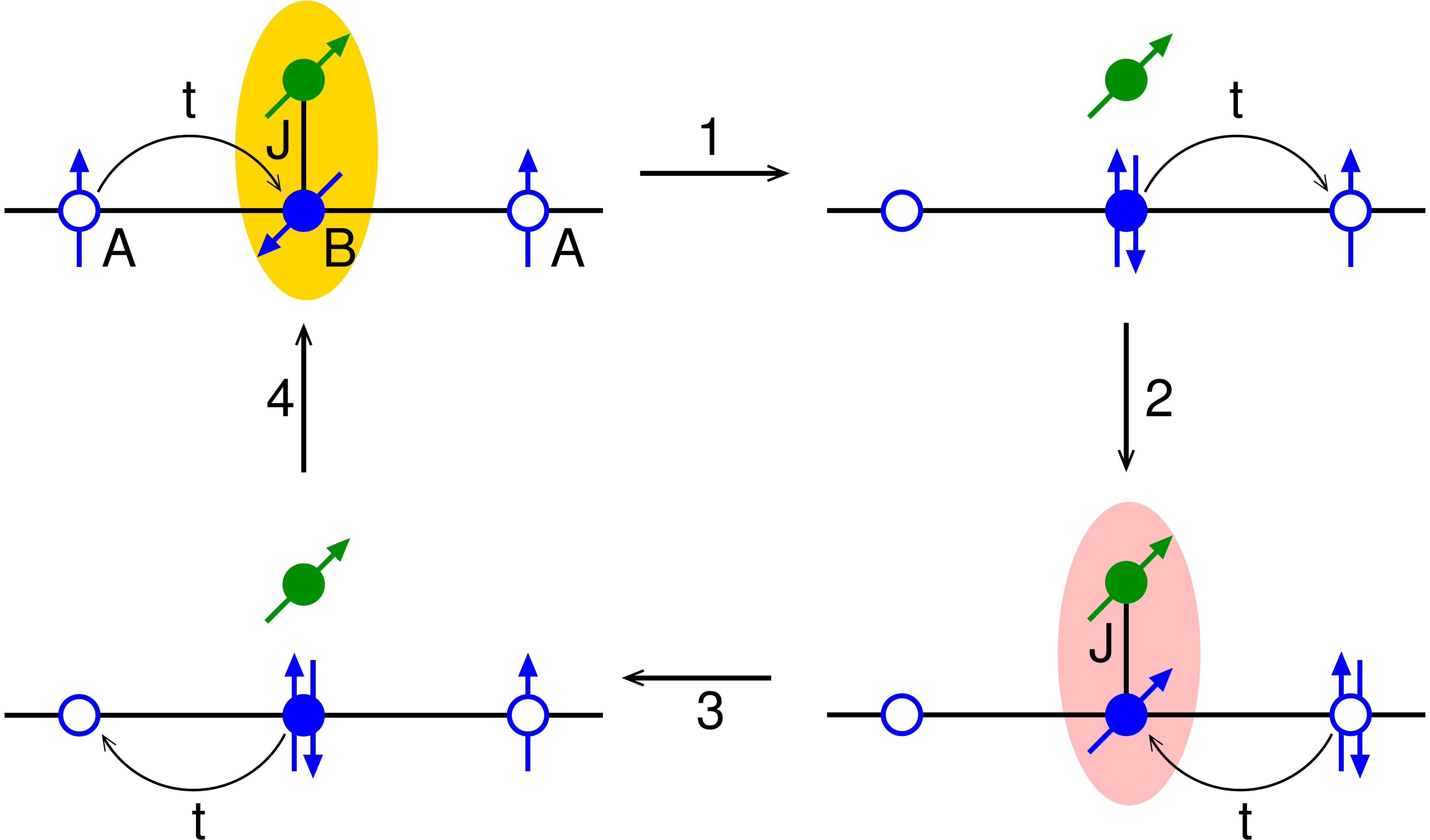}  }
\caption{(Color online) Illustration of a typical fourth-order process consisting of the following four steps:
(1) A Kondo singlet at site $i \in B$ is broken, i.e.\ it is excited to a state where the site $i$ is either empty or doubly occupied.
(2) The system is further excited to a local triplet state at the same site. 
(3) The system returns to a broken Kondo singlet state.
(4) The local Kondo singlet is restored at site $i$.}
\label{fig:4thorder}
\end{figure}

As for any finite term in Eq.\ \eqref{perturbation_0} a local Kondo singlet must be broken and reconstructed again, an even number of hopping processes is required, i.e.\ all odd-order terms vanish.
Hence, the first non-vanishing term is of second order.
It comprises processes where, due to virtual hopping, one of the Kondo singlets is excited to a broken Kondo-singlet state, i.e.\ to a state where the corresponding $B$-sublattice site is either empty or doubly occupied. 
After a second hopping process the Kondo singlet is restored. 
The calculation shows, however, that due to the particle-hole symmetry of the unperturbed Hamiltonian ${\cal H}_0$ (see Eq.\ \eqref{H0_Kondo}), i.e., of the individual Kondo singlets, such processes essentially cancel each other. 
Let us stress that this does not constrain the filling for the full problem given by the Hamiltonian $\cal H$ (see Eq.\ \eqref{Hamiltonian_Kondo}).
The second-order term turns out to be merely given by a constant
\begin{equation}
\label{perturbation_2_Kondo} 
t^2 \sum_{i \in B} \frac{P_0 {\tilde{\cal H}}_1^{(i)} P_1^{(i)} {\tilde{\cal H}}_1^{(i)} P_0}{E_0-E_1}=P_0\frac{4t^2}{3J}\sum_{i \in B} Z_{i}   \, ,
\end{equation}
where $Z_{i}$ is the coordination number of lattice site $i$. 
Consequently, second-order perturbation theory cannot lift the ground-state degeneracy of ${\cal H}_0$. 

This is achieved, however, with the fourth-order contributions. 
Those contain three different types of processes. 
Two of them involve the excitation of two different Kondo singlets to broken Kondo-singlet states and differ in the order in which Kondo singlets are restored. 
The calculation shows that, similar to the second-order term, due to the particle-hole symmetry of ${\cal H}_0$, these two types of processes give rise to an essentially irrelevant constant only:
\begin{eqnarray}
\label{perturbation_4_2Kondo}
&&
t^4 \sum_{i_1, i_2 \in B}^{i_1 \not= i_2}
\frac{P_0 {\tilde{\cal H}}_1^{(i_1)} P_{1,0}^{(i_1,i_2)}{\tilde{\cal H}}_1^{(i_2)}P_{1,1}^{(i_1,i_2)} {\tilde{\cal H}}_1^{(i_1)} P_{0,1}^{(i_1,i_2)} {\tilde{\cal H}}_1^{(i_2)} P_0}{(E_0-E_1)^2(E_0-E_3)} 
\nonumber \\
&+&
t^4  \sum_{i_1, i_2 \in B}^{i_1 \not= i_2}
\frac{P_0 {\tilde{\cal H}}_1^{(i_1)} P_{1,0}^{(i_1,i_2)}{\tilde{\cal H}}_1^{(i_2)}P_{1,1}^{(i_1,i_2)} {\tilde{\cal H}}_1^{(i_2)} P_{1,0}^{(i_1,i_2)} {\tilde{\cal H}}_1^{(i_1)} P_0}{(E_0-E_1)^2(E_0-E_3)}  
\nonumber \\
&=&
- P_0\frac{16t^4}{27J^3} \sum_{i_1, i_2 \in B}^{i_1 \not= i_2} Z_{i_1i_2} 
\, .
\end{eqnarray}
Here $Z_{i_1i_2}$ is the number of sites which are neighbors of both, site $i_1$ and site $i_2$. 
We also use the notation $P_{\alpha,\beta}^{(i_1,i_2)}= P_\alpha^{(i_1)} P_\beta^{(i_2)}$. 

The third type of processes consists of a twofold excitation of the same Kondo singlet, first to a broken Kondo-singlet state, and later to a triplet state, see Fig.~\ref{fig:4thorder} for an example. 
These processes are decisive for the effective Hamiltonian:
\begin{eqnarray}
{\cal H}_{\rm eff}=
t^4 \sum_{i \in B} \frac{P_0 {\tilde{\cal H}}_1^{(i)} P_1^{(i)} {\tilde{\cal H}}_1^{(i)} P_2^{(i)} {\tilde{\cal H}}_1^{(i)} P_1^{(i)} {\tilde{\cal H}}_1^{(i)}P_0}{(E_0-E_1)^2(E_0-E_2)}  \, .
\end{eqnarray}
Inserting the representations of the projection operators, Eqs.\ \eqref{P1} and \eqref{P2}, and evaluating the corresponding matrix elements, we obtain the following explicit form for the effective low-energy Hamiltonian:
\begin{equation}
\label{Heff_0} 
{\cal H}_{\rm eff}=-P_0\frac{\alpha_K}{4}\sum_{i \in B}\sum_{j_1,\ldots,j_4 \in A_i} \sum_\sigma c_{j_1,\sigma}^\dagger c_{j_2,-\sigma}c_{j_3,-\sigma}^\dagger c_{j_4,\sigma} \: .
\end{equation}
Here 
\begin{equation}
\label{alpha_Kondo} 
\alpha_K=64t^4/3J^3
\end{equation}
is the effective coupling constant. 
Note that there is a single energy scale only.

\section{Strong coupling perturbation theory for the depleted Anderson lattice model}
\label{sec:and}

The depleted Anderson lattice model can be treated in a very similar way.
Here, the spin-$1/2$ Kondo impurities are replaced by Anderson impurities, i.e.\ by correlated sites with Hubbard interaction $U$ which are coupled to the conduction-electron system via a local hybridization term with hybridization constant $V$.  
The two models with Anderson and with Kondo impurities, respectively, can be mapped onto each other by means of the Schrieffer-Wolff transformation \cite{SW66,SN02} for all finite $J=8V^2/U$ in the limit $V \rightarrow \infty$ and $U \rightarrow \infty$. 
If these conditions are satisfied and if $J\gg |t|$, both models are obviously described by the same strong-coupling effective Hamiltonian. 
The open question is what happens away from the Kondo limit when there is no direct mapping.

To investigate this issue we perform perturbation theory for the depleted Anderson model as well. 
Now the unperturbed part of the Hamiltonian, ${\cal H}_0^{(i)}$, replacing Eq. \eqref{H0_Kondo}, is given by
\begin{equation}
\label{H0_Anderson}
 {\cal H}_0^{(i)}=V \sum_\sigma \left(c_{i,\sigma}^\dagger f_{i,\sigma} + \mbox{H.c.}\right)+U\left(n_{i,\uparrow}^f-\frac{1}{2}\right)\left(n_{i,\downarrow}^f-\frac{1}{2}\right)
\end{equation}
while the perturbation ${\cal H}_1$ is unchanged and still given by Eq.\ \eqref{H1}. 
The strong-coupling limit in the Anderson case corresponds to the limit $|t| \ll V$. 
The subsequent derivation is again valid for total particle numbers in the range $R < N <2L-R$. 

As compared to the Kondo case, the excitation spectrum of the unperturbed Hamiltonian is much richer for the depleted Anderson lattice. 
Still the unperturbed problem at a site $i$ of the sublattice $B$, given by ${\cal H }_0^{(i)}$, can be solved analytically, see Ref.\ \onlinecite{Lan98} and Appendix \ref{App_Unperturbed_Hamiltonian} for the resulting eigenvectors $|i;q,m,l\rangle$ and eigenvalues ${\cal E}_{q,m,l}^{(i)}$.
Here $q=0,\ldots  4$ and $m=0, \pm 1/2, \pm 1$ denote the total particle number and the magnetic quantum number of the corresponding eigenstate of the two-site problem ${\cal H }_0^{(i)}$ and are conserved quantum numbers. 
Furthermore, $l$ enumerates the orthogonal states in a sector with fixed $q$ and $m$.

The ground-state energy of the unperturbed Hamiltonian ${\cal H}_0$ is $E_0=-R\frac{1}{4}\sqrt{U^2+64V^2}$, and the respective projection operator is
\begin{equation}
\label{P0_Anderson}
P_0=\bigotimes_{i \in B} P_0^{(i)}  \quad {\rm with} \quad  P_0^{(i)}=|i;2,0,0\rangle \langle i;2,0,0|  \, .
\end{equation}

The excitations of lowest energy result from virtual hopping processes, in which an electron is temporarily added or removed from one of the Anderson singlets. 
The corresponding excitation energy is $E_1-E_0=\frac{1}{4}\sqrt{U^2+64V^2}-\frac{1}{4}\sqrt{U^2+16V^2}$. 
Perturbation theory is justified if $|t| \ll E_1-E_0$. 
This condition is fulfilled for $|t| \ll V$ and only weakly depends on $U$.  
The respective projection operator is given by Eq.\ \eqref{P1} but with
\begin{eqnarray}
\label{P1_Anderson} 
P_1^{(i)}= \sum_{q=1,3}\sum_{m=\pm 1/2} |i;q,m,0\rangle\langle i; q,m,0|  \, 
\end{eqnarray}
in the case of Anderson impurities.

The second excited energy level $E_2=E_0+\frac{1}{4}\sqrt{U^2+64V^2}-\frac{1}{4}U$ corresponds to spin triplet states. 
The projection operator $P_2^{(i)}$ is given by Eq. \eqref{P2}, but with 
$|T^0_i\rangle =|i;2,0,1\rangle$ and $|T^{\pm}_i\rangle = |i;2,\pm 1,0\rangle$.

The third excited energy level, similar to the Kondo case, contains two broken singlet states at different impurities and its energy is $E_3=E_0+2\left(\frac{1}{4}\sqrt{U^2+64V^2}-\frac{1}{4}\sqrt{U^2+16V^2}\right)$. The corresponding 
projection operator is specified by Eqs.~\eqref{P3} and \eqref{P1_Anderson}.

Here, we have to consider also higher excitations which have no analogue in the Kondo case. 
The fourth excited energy level corresponds to the isospin triplet states 
$|i;0,0,0\rangle$, $|i;2,0,2\rangle$, $|i;4,0,0\rangle$ (see also  Eq.\ \eqref{isospin}), i.e.\ states where either one of the Anderson singlets is excited by adding or removing two electrons with opposite spin, or it corresponds to the second excited state in the sector $q=2$ and $m=0$. 
The energy of this excitation is $E_4-E_0=\frac{1}{4}\sqrt{U^2+64V^2}+\frac{1}{4}U$, and its projection operator is given by 
\begin{eqnarray}
P_4
&=&
\sum_{i \in B} P_4^{(i)} \bigotimes_{i' \in B}^{ i'\not= i} P_0^{(i')}  \quad {\rm with}  
\nonumber \\
P_4^{(i)}
&=& 
\sum_{q=0,4} |i;q,0,0\rangle\langle i;q,0,0|+|i;2,0,2\rangle\langle i;2,0,2|  \, . 
\nonumber \\ 
\end{eqnarray}

The fifth excited energy level refers to excited states which are also obtained by adding or removing an electron from the Anderson singlet. 
It is given by $E_5=E_0+\frac{1}{4}\sqrt{U^2+64V^2}+\frac{1}{4}\sqrt{U^2+16V^2}$, and the corresponding projection operator reads
\begin{eqnarray}
P_5
&=&
\sum_{i \in B} P_5^{(i)} \bigotimes_{i' \in B}^{ i'\not= i} P_0^{(i')}  \quad {\rm with}  
\nonumber \\
P_5^{(i)}
&=& 
\sum_{q=1,3}\sum_{m=\pm 1/2} |i;q,m,1\rangle\langle i; q,m,1| 
\, . 
\end{eqnarray}

The sixth excited level is reached by either exciting an Anderson singlet to the highest energy state in the sector $q=2$ and $m=0$ or by breaking two Anderson singlets with different excitation energies. 
The corresponding energy is $E_6=E_0+\frac{1}{2}\sqrt{U^2+64V^2}$ and
\begin{eqnarray}
P_6
&=&
\sum_{i \in B} P_6^{(i)} \bigotimes_{i' \in B}^{ i'\not= i} P_0^{(i')}+ \sum_{i_1, i_2 \in B}^{i_1 \ne i_2}  P_1^{(i_1)} P_4^{(i_2)} \bigotimes_{i' \in B}^{i'\not=i_1, i_2} P_0^{(i')} 
\nonumber \\
&& 
{\rm with} \quad  P_6^{(i)}=|i;2,0,3\rangle\langle i; 2,0,3|  \, .
\end{eqnarray}

The excited state of highest energy which is needed for fourth-order perturbation theory involves two broken Anderson singlets with highest excitation energy. 
We find
$E_7=E_0+2\left(\frac{1}{4}\sqrt{U^2+64V^2}+\frac{1}{4}\sqrt{U^2+16V^2}\right)$ and 
\begin{eqnarray}
\label{P7_Anderson}
P_7= \sum_{i_1, i_2 \in B}^{i_1 > i_2}  P_4^{(i_1)} P_4^{(i_2)} \bigotimes_{i' \in B}^{i'\not=i_1, i_2} P_0^{(i')} \: .
\end{eqnarray}

With the low-lying excitations of the unperturbed depleted Anderson lattice and with the above projection operators at hand, we can easily derive the effective Hamiltonian. 
As in the Kondo case and for the same reasons, fourth-order perturbation theory is necessary to lift the ground-state degeneracy of ${\cal H}_0$.
Furthermore, odd-order terms of the perturbation theory are vanishing, and the second-order terms, due to the particle-hole symmetry of ${\cal H}_0^{(i)}$, provide us with a constant contribution only:
\begin{equation}
\label{perturbation_2_Anderson} 
t^2 \sum_{i \in B} \sum_{l=1,5} \frac{P_0 {\tilde{\cal H}}_1^{(i)} P_l^{(i)} {\tilde{\cal H}}_1^{(i)} P_0}{E_0-E_l}
=
- P_0 \frac{t^2\left(U^2+48V^2\right)  \sum\limits_{i \in B} Z_{i}}{12V^2\sqrt{U^2+64V^2}}   \, .
\end{equation}
At fourth order, perturbation theory again involves three different processes. 
Two of them are given by excitations of two different Anderson singlets while the third one consists in a double excitation of an Anderson singlet. 
The first two processes, due to the particle-hole symmetry of ${\cal H}_0^{(i)}$, result in the constant
\begin{eqnarray}
\label{perturbation_4_2Anderson}
&&
t^4 \sum_{i_1, i_2 \in B}^{i_1 \not= i_2} \sum_{l_1=1,5} \sum_{l_2=1,5} \frac{1}{(E_0-E_{l_1})(E_0-E_{l_1l_2})(E_0-E_{l_2})} 
\nonumber \\
& \times &
\left( 
P_0 {\tilde{\cal H}}_1^{(i_1)} P_{l_1,0}^{(i_1,i_2)}{\tilde{\cal H}}_1^{(i_2)}P_{l_1,l_2}^{(i_1,i_2)} {\tilde{\cal H}}_1^{(i_1)} P_{0,l_2}^{(i_1,i_2)} {\tilde{\cal H}}_1^{(i_2)} P_0 
\right. 
\nonumber \\
&+& 
\left. 
P_0 {\tilde{\cal H}}_1^{(i_1)} P_{l_1,0}^{(i_1,i_2)}{\tilde{\cal H}}_1^{(i_2)}P_{l_1,l_2}^{(i_1,i_2)} {\tilde{\cal H}}_1^{(i_2)} P_{l_1,0}^{(i_1,i_2)} {\tilde{\cal H}}_1^{(i_1)} P_0
\right) 
\nonumber \\
& = &
- t^4 \frac{U^6 + 148 U^4 V^2 + 7056 U^2 V^4 + 110592 V^6}{864V^6(U^2+64V^2)^{3/2}} \sum_{i_1, i_2 \in B}^{i_1\not=i_2}Z_{i_1i_2} 
\nonumber \\
&-&
\frac{2U^2 t^4}{9V^2 (U^2+64V^2)^{3/2}}\sum_{i_1, i_2 \in B}^{i_1\not=i_2}Z_{i_1}Z_{i_2} 
\, .
\end{eqnarray}
Here $P_{l_1,l_2}^{(i_1,i_2)} = P_{l_1}^{(i_1)} P_{l_2}^{(i_2)}$, and $E_{l_1l_2}$ is the energy of the excited state with two broken singlets.
Correspondingly, $E_{1,1}=E_3$, $E_{1,5}=E_{5,1}=E_6$ and $E_{5,5}=E_7$.

Finally, the third type of processes do remove the ground-state degeneracy of the unperturbed Hamiltonian ${\cal H}_0$. 
We have
\begin{eqnarray}
{\cal H}_{\rm eff}
&=&
t^4 \sum_{i \in B}\sum_{l_1=1,5}\sum_{l_2=2,4,6}\sum_{l_3=1,5}{\cal H}_{\rm eff}^{(i)}
\nonumber \\
{\cal H}_{\rm eff}^{(i)}
&=&
\frac{P_0 {\tilde{\cal H}}_1^{(i)} P_{l_1}^{(i)} {\tilde{\cal H}}_1^{(i)} P_{l_2}^{(i)} {\tilde{\cal H}}_1^{(i)} P_{l_3}^{(i)} {\tilde{\cal H}}_1^{(i)}P_0}{(E_0-E_{l_1})(E_0-E_{l_2})(E_0-E_{l_3})}  
\, .
\end{eqnarray}
This yields exactly the same effective Hamiltonian as for the depleted Kondo lattice model, i.e.\ Eq.\ \eqref{Heff_0}, but now the coupling constant $\alpha$ depends on $U$ and $V$ as follows:
\begin{eqnarray}
\label{alpha_Anderson} 
\alpha_A=t^4 \frac{U^3+48UV^2}{24V^6} \, .
\end{eqnarray}
Opposed to the Kondo case, the third type of processes result in an additional constant term:
\begin{eqnarray}
&-&
32t^4\left(\frac{2V^2(U^4 + 116 U^2 V^2 + 3456 V^4)}
{9 V^4 (U^2 + 64 V^2)^{3/2} \left(U +\sqrt{U^2 + 64 V^2} \right)^2 }  
\right.
\nonumber \\
&+&
\left. 
\frac{U\left(U^2+72V^2\right) }
{9 V^2 (U^2 + 64 V^2) \left(U +\sqrt{U^2 + 64 V^2} \right)^2 }\right) \sum_{i\in B}Z_i^2 \: .
\nonumber \\
\end{eqnarray}
This term is vanishing in the Kondo limit, i.e.\ for $U \rightarrow \infty$, $V \rightarrow \infty$ but $V^2/U \to \mbox{const.}$ as it should be the case.

A reason why we get exactly the same Hamiltonians, apart from the coupling constants $\alpha_K$ and $\alpha_A$, is that the perturbation theories performed for those two systems must generate effective interactions on the nearest-neighbor sites of the Kondo singlets $i\in B$ which are highly constrained by U(1) particle number, the SU(2) spin and the SU(2) isospin symmetry of the original unperturbed Hamiltonians.

\section{Comparing depleted Anderson and Kondo lattices}
\label{sec:comp}

The two expressions for the effective coupling constants, Eqs.\ (\ref{alpha_Kondo}) and (\ref{alpha_Anderson}), become identical when the two models can be mapped onto each other, i.e.\ in the limit $V \rightarrow \infty$ and $U \rightarrow \infty$ but $J=8V^2/U \gg |t|$.
This had to be expected from the Schrieffer-Wolff transformation. \cite{SW66,SN02}

For finite but large on-site interaction ($U \gg V \gg |t| $), i.e.\ when charge fluctuations at the impurity sites are strongly suppressed but non-zero, one can still (at least formally) introduce an effective coupling $J=8V^2/U$ between impurity and conduction-electron spins. 
With this, Eq.\ \eqref{alpha_Anderson} reads 
\begin{eqnarray}
\label{alpha_Anderson_Kondo_limit} 
\alpha_A= \frac{64t^4}{3J^3}\left(1+\frac{48V^2}{U^2}\right) \, . 
\end{eqnarray}
Thereby it becomes obvious that only for $U \rightarrow \infty$ the coupling constants are equal, $\alpha_A=\alpha_K$, while for any finite but large $U$, the coupling $\alpha_A$ in the Anderson case is larger as compared to the coupling $\alpha_K$ in the Kondo case for the same value of $J$. 
This will e.g.\ affect finite-temperature properties and critical temperatures: 
For the same $J$, a phase transition must take place at a higher temperature in the strong-coupling limit of the depleted Anderson model as compared to the Kondo case.
This is interesting as the opposite might have been expected because of the additional charge degrees of freedom and the related charge fluctuations in the Anderson case.

It is worth pointing out that the first excitation gap of the two-site problem with an Anderson impurity, namely $\Delta{\cal E}_1^A= \frac{1}{4}\sqrt{U^2+64V^2}-\frac{1}{4}\sqrt{U^2+16V^2}$ is smaller, for any $U$ and $V$, than the corresponding gap for a Kondo impurity which is given by $\Delta{\cal E}_1^K=\frac{3}{4}J$ at $J=8V^2/U$. 
Therefore, to satisfy the condition for perturbation theory to be reliable, $\Delta{\cal E}_1 \gg t$, one needs a stronger coupling $8V^2/U$ for the Anderson case as compared to $J$ in the Kondo case.

\begin{figure}[t]
\centerline{\includegraphics[width=0.45\textwidth]{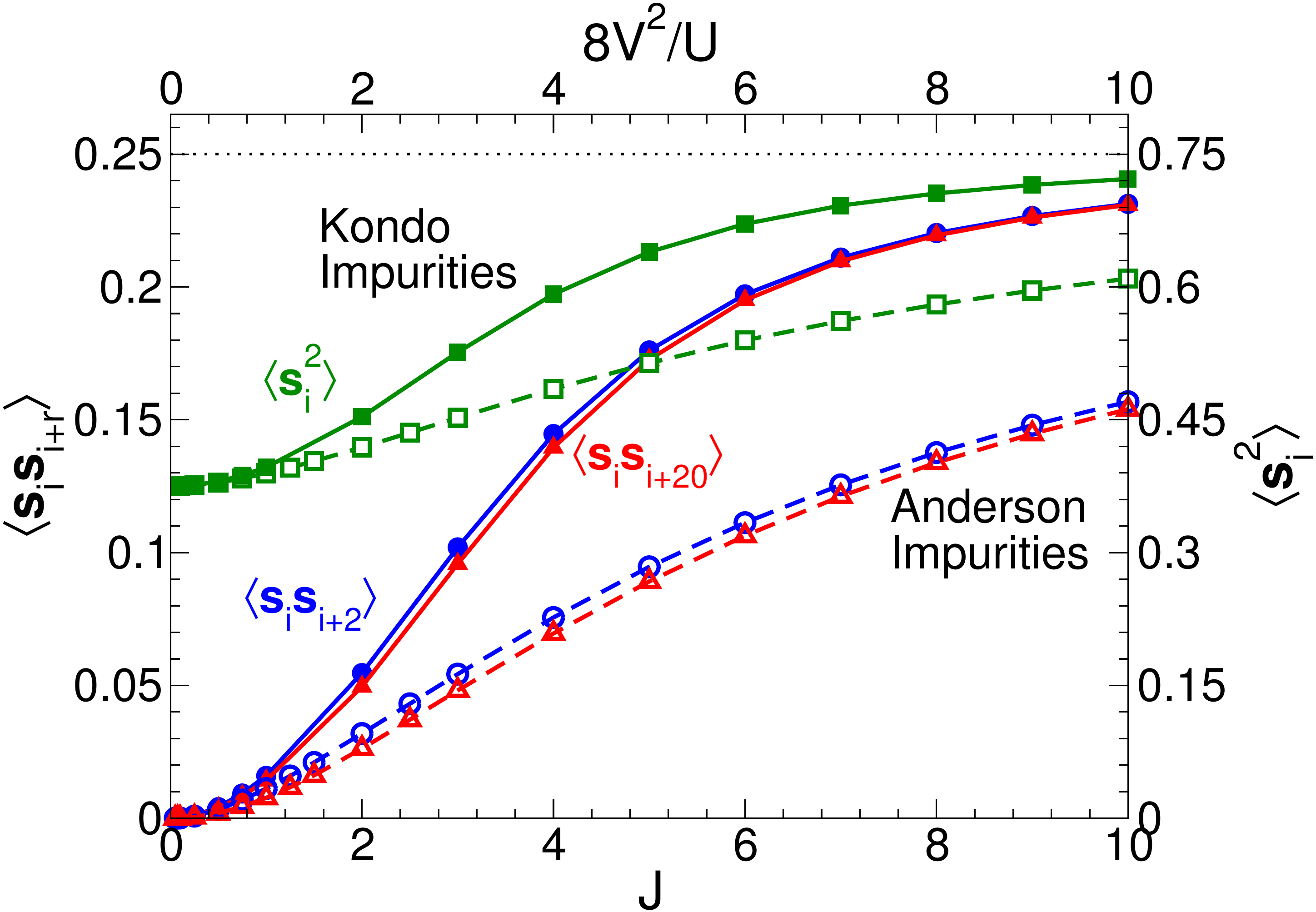}}
\caption{(Color online)
Density-matrix renormalization group (DMRG) calculations for the local moment $\langle {\bf s}_i^2 \rangle$ on the $A$-sublattice sites and for the spin-spin correlation $\langle {\bf s}_i \, {\bf s}_{i+r} \rangle$ between the $A$-sublattice sites (with $r = 2, 20$ and the central site $i=24$).
Calculations are performed for a system with $L=49$ uncorrelated conduction-electron sites ($A$ and $B$) and $R=25$ impurities, see Fig.\ \ref{fig:depleted_1D} for the system geometry. 
Solid lines, filled symbols: spin-1/2 Kondo impurities.
Dashed lines, open symbols: Anderson impurities.
Results are shown as functions of $J$ and of $8V^{2}/U$ (with Hubbard $U=8$), respectively, and cover the crossover from the RKKY regime at weak coupling to the strong-coupling regime ($J \gg |t|$ and $V \gg |t|$).
The energy scale is fixed by $|t|=1$.
}
\label{fig:correction}
\end{figure}

We have checked this by numerical calculations using the density-matrix renormalization group (DMRG) \cite{Whi92,Sch11} for the one-dimensional geometry sketched in Fig.\ \ref{fig:depleted_1D}.  
A standard implementation based on matrix-product states and matrix-product operators is employed (see Ref.\ \onlinecite{TSRP12} for some details). 
Previous studies have demonstrated that the system has a ferromagnetic ground state at half-filling. \cite{STP13,TSP14}
Here, to compare the depleted Anderson and Kondo lattice with each other, we compute the local moment $\langle {\bf s}_{i}^2 \rangle$ as well as the spin-spin correlation $\langle {\bf s}_{i}  {\bf s}_{i'} \rangle$ for different $A$-sublattice sites $i,i'$, see Fig.\ \ref{fig:correction}.

The results are qualitatively similar for both models:
In the RKKY regime, i.e.\ for weak coupling ($J \ll |t|$ and $V \ll |t|$) the local moments on the $A$-sublattice sites are delocalized, $\langle {\bf s}_{i}^2 \rangle = 3/8$, and spins on different $A$-sublattice sites are almost uncorrelated 
$\langle {\bf s}_{i}  {\bf s}_{i'} \rangle \simeq 0$. 
Contrary, the impurity spins are strongly correlated $\langle {\bf S}_{i}  {\bf S}_{i'} \rangle \simeq 1/4$ (not shown). 
With increasing of $J$, the local moment $\langle {\bf s}_{i}^2 \rangle$ and the spin-spin correlations
$\langle {\bf s}_{i}   {\bf s}_{i'} \rangle$ are increasing, and in the strong-coupling limit ($J \gg |t|$ and $V \gg |t|$) approach their limiting values $\langle {\bf s}_{i}^2 \rangle=3/4$ and $\langle {\bf s}_{i}   {\bf s}_{i'} \rangle=1/4$. 
The spin correlations are ferromagnetic and only very weakly depend on the distance between spins as it is characteristic for the symmetry-broken ferromagnetic ground state 
(see Ref.\ \onlinecite{TSP14} for a detailed discussion of the inverse indirect magnetic exchange mechanism which governs the system's magnetic properties in the strong-coupling limit).

Here, we like to stress that, comparing the results obtained for Kondo and Anderson impurities, convergence to the strong-coupling limit is considerably faster for the Kondo model if plotted as functions of $J$ and $8V^2/U$, respectively. 
This nicely confirms the above-mentioned condition for the validity of strong-coupling perturbation theory based on the excitation gap in the two-site problems with Kondo and Anderson impurities.

For the depleted Anderson lattice this condition is $\frac{1}{4}\sqrt{U^2+64V^2}-\frac{1}{4}\sqrt{U^2+16V^2} \gg t$. 
Therefore, the effective model should not only apply in the Kondo limit and for strong $8V^{2}/U$, but also in the limit when $U \ll V$ but $V \gg t$. 
The latter also includes the non-interacting system $U=0$. 
However, according to Eq.\ \eqref{alpha_Anderson} the coupling constant $\alpha_A=0$ in this case. 
This indicates that perturbation theory does not lift the ground-state degeneracy of the unperturbed system up to fourth order. 
This is due to the fact that for $U=0$ spin triplet states ($|2,0,1\rangle$, $|2,\pm 1,0\rangle$) and triplet isospin states 
($|0,0,0\rangle$, $|2,0,2\rangle$, $|4,0,0\rangle$, see also Eq. \eqref{isospin}) have the same energy. 
In fact, this degeneracy cannot be lifted in any order. 
This is due to the fact that for $U=0$ the system under consideration has a flat band dispersion \cite{TSP14} and thus a highly degenerate ground state for any $V$. 

\section{Spin-isospin representation}
\label{sec:spiniso}

To discuss the physics of the effective Hamiltonian Eq.\ \eqref{Heff_0} we rewrite it in two different ways, starting with a representation in terms of local spins and local isospins which is possible for a bipartite lattice.
To this end, we introduce the local isospin at a site $j$ as
\begin{eqnarray}
\label{isospin}
{\bf t}_j =\frac{1}{2} \left(c_{j,\uparrow}^\dagger, (-1)^j c_{j,\downarrow}^{\phantom\dagger}\right) \cdot {\boldsymbol \sigma} \cdot \left(c_{j,\uparrow}^{\phantom\dagger}, (-1)^j c_{j,\downarrow}^\dagger \right)^{T} \, .
\end{eqnarray}
In an eigenstate of $t^z_j$ with eigenvalue $m_t^{(j)}=-1/2$ the site $j$ is unoccupied, while it is doubly occupied for $m_t^{(j)}=1/2$.
After straightforward calculations we find
\begin{equation}
{\cal H}_{\rm eff} = \alpha \sum_{i\in B}\tilde{\cal H}_{\rm eff}^{(i)}
\end{equation}
with 
\begin{eqnarray}
\label{Heff_fin_gen}
\tilde{\cal H}_{\rm eff}^{(i)}
&=&
- \sum_{j_1,j_2 \in A_i}^{j_1>j_2} \left({\bf s}_{j_1} {\bf s}_{j_2} - {\bf t}_{j_1}  {\bf t}_{j_2}  \right) 
\nonumber \\
&+&
\frac{1}{2}\sum_{j \in A_i} \left(n_{j,\uparrow}-\frac{1}{2}\right) \left( n_{j,\downarrow}-\frac{1}{2}\right)  
\nonumber\\
&-&
\frac{1}{2}\sum_{j_1,j_2 \in A_i} \sum_\sigma c_{j_1,\sigma}^\dagger c_{j_2,\sigma}^{\phantom\dagger}
\left( \frac{1}{2}Z_{i} - \sum_{j\in A_{i}} n_{j,-\sigma}  \right)  
\nonumber\\
&+&
\frac{1}{2}\sum_{j_1, j_2, j_3 \in A_i}^{\rm all~different} (c_{j_1,\uparrow}^\dagger c_{j_2,\uparrow}+\mbox{H.c.})
(c_{j_1,\downarrow}^\dagger c_{j_3,\downarrow}+\mbox{H.c.})
\nonumber\\
&+&
\frac{1}{2}\sum_{j_1, j_2, j_3, j_4 \in A_i}^{\rm all~different}
c_{j_1,\uparrow}^\dagger c_{j_2,\downarrow}^\dagger c_{j_3,\downarrow}^{\phantom\dagger}  c_{j_4,\uparrow}^{\phantom\dagger}  \, .
\end{eqnarray}
For simplicity, here and from now on we suppress the projection operator $P_0$ in the notation and assume that all sites of the $B$-sublattice are occupied by a single electron forming a singlet state with the impurity.

The first term in the Hamiltonian Eq.\ \eqref{Heff_fin_gen} is a Heisenberg-type ferromagnetic spin interaction and is responsible for the ferromagnetic order found in Refs.\ \onlinecite{STP13,TSP14}. 
The second term describes an antiferromagnetic isospin interaction which favors a charge-density wave or $\eta$-superconductivity.\cite{TSU97b}
However, the repulsive local Hubbard interaction, i.e.\ the third term in Eq.\ \eqref{Heff_fin_gen}, suppresses the formation of isospins and rather supports the formation of spins. 
The fourth term describes a correlated hopping across the $i$-th Kondo singlet where hopping of electrons with spin $\sigma$ depends on the number of electrons with spin $-\sigma$ on the neighboring sites of $i$-th Kondo singlet. 
The last two terms in the Hamiltonian Eq.~\eqref{Heff_fin_gen} correspond to non-local pair-hopping processes of a spin-up and a spin-down electron in the vicinity of the same Kondo singlet. 
For the first, the two pair-hopping processes share one $A$-sublattice site, while for the second all $A$-sublattice sites involved are mutually different.

It is obvious that the last two terms only exist for two and higher dimensions. 
For a one-dimensional system we have
\begin{eqnarray}
\label{Heff_fin_1D}
  {\cal H}_{\rm eff}/\alpha
&=&
  - \sum_{\langle j_1,j_2 \rangle \in A}^{j_1>j_2}
  \left({\bf s}_{j_1} {\bf s}_{j_2} - {\bf t}_{j_1}  {\bf t}_{j_2}  \right) 
  \nonumber \\
&+& 
  \sum_{j \in A} \left( n_{j,\uparrow} - \frac{1}{2} \right) 
                        \left( n_{j,\downarrow} - \frac{1}{2} \right)  
  \nonumber\\
&-&
  \frac{1}{2} \sum_{\langle i,j \rangle \in {\rm A}}\sum_\sigma c_{i,\sigma}^\dagger c_{j,\sigma}  \left(1-n_{i,-\sigma}-n_{j,-\sigma}\right)   \, ,
  \nonumber\\
\end{eqnarray}
where $\langle i,j \rangle$ means summation over neighboring $A$-sublattice sites. 
This has been discussed extensively in Ref.\ \onlinecite{STP13}. 

\section{Non-local spin representation}
\label{sec:nonlocal}

To analyze the ground-state properties of the effective Hamiltonian, it is instructive to rewrite it in a different way. 
For simplicity, we consider a system with periodic boundary conditions and translational symmetry such that the coordination number $Z_i=Z$ is a constant. 

\begin{figure}[bpt]
\centerline{\includegraphics[width=0.4\textwidth]{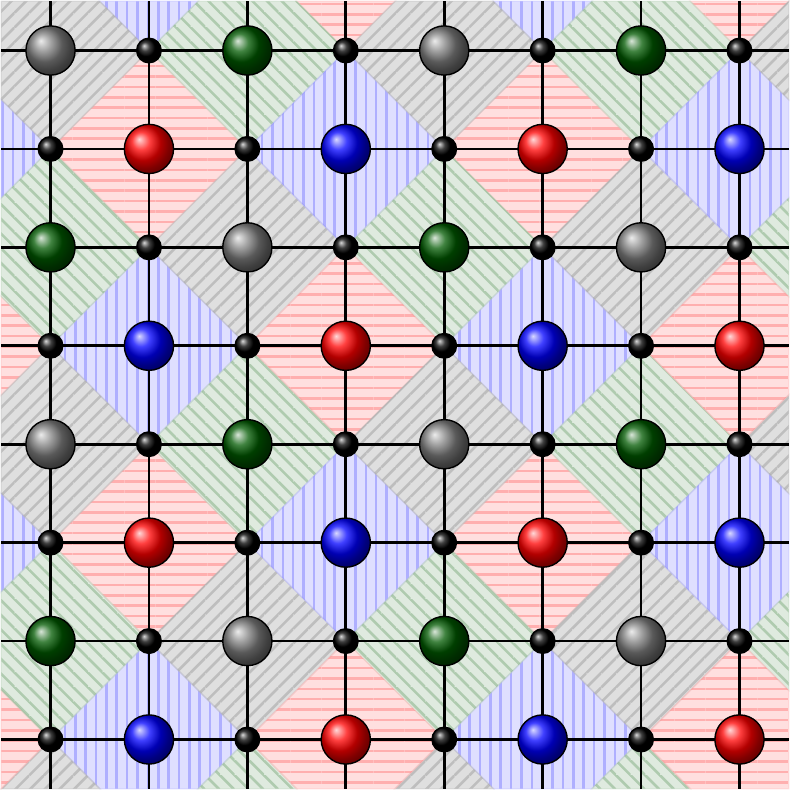}  }
\caption{(Color online) 
Schematic illustration of the effective Hamiltonian ${\cal H}_{\rm eff}$ (see main text) for a square lattice. 
Small black circles correspond to $A$-sublattice sites.
Large circles correspond to $B$-sublattice sites. 
For each $B$-sublattice site $i$, we introduce a spin ${\bm{\mathcal S}}_{i,1}$ which operates on electronic basis states at the neighboring $A$-sublattice sites. 
These sites form a plaquette centered around site $i$. 
Due to their non-local character, spins ${\bm{\mathcal S}}_{i,1}$ centered at neighboring $B$-sublattice sites do not commute.
The $B$-sublattice is thus subdivided into $Z=4$ different and non-overlapping groups of sites (forming sub-sublattices) represented by four different colors (different patterns). 
Pairs of spins belonging to the same group are commutative.
Note that each group fully covers the $A$-sublattice.
}
\label{fig:plaquette}
\end{figure}

We divide the $B$-sublattice into $Z$ sub-sublattices $B_m$ which we refer to as groups. 
Fig.~\ref{fig:plaquette} illustrates the tiling of the $B$ sublattice for the case of a two-dimensional lattice.
Each group $B_{m}$ is represented by a different color.
The effective Hamiltonian, Eq.\ \eqref{Heff_0}, is given by
\begin{eqnarray}
\label{Effective_Hamiltonian_final_1}
{\cal H}_{\rm eff}=\sum_{m=1}^Z {\cal H}_{\rm eff}^{(m)} \, ,
\end{eqnarray}
where
\begin{eqnarray}
\label{Effective_Hamiltonian_m_0}
{\cal H}_{\rm eff}^{(m)}=\alpha \sum_{i \in B_m} \tilde{\cal H}_{\rm eff}^{(i)}
\end{eqnarray}
with $\alpha=\alpha_{K}$ or $\alpha=\alpha_{A}$
and where the different terms 
\begin{eqnarray}
\label{Effective_Hamiltonian_i_0}
\tilde{\cal H}_{\rm eff}^{(i)}
=
-\frac{1}{4} \sum_{j_1,\ldots,j_4 \in A_i} \sum_\sigma c_{j_1,\sigma}^\dagger c_{j_2,-\sigma}c_{j_3,-\sigma}^\dagger c_{j_4,\sigma}
\end{eqnarray}
are pairwise commutative for all $i$ in the same group $B_m$.
Namely, for a given $m$ the Hamiltonian ${\cal H}_{\rm eff}^{(m)}$ operates on the sites of the sublattice $A$, and furthermore
each term $\tilde{\cal H}_{\rm eff}^{(i)}$ operates on a different plaquette of $A$-sublattice sites only. 
(For the $D=1$ case, each $\tilde{\cal H}_{\rm eff}^{(i)}$ operates on a different {\em bond} of two $A$-sublattice sites).
These plaquettes (bonds) are centered around the sites $i \in B_{m}$.
This also implies that we have $Z$ different tilings of the sublattice $A$, each specified by $m$. 
Each of the tilings of sublattice $A$ covers the whole sublattice $A$ (see a set of plaquettes of the same color in Fig.\ \ref{fig:plaquette}).
Therefore, it is obvious that the problem specified by ${\cal H}_{\rm eff}^{(m)}$ for a given $m$ becomes exactly solvable.
However, the different ${\cal H}_{\rm eff}^{(m)}$ for different $m=1,...,Z$ do not commute.
This makes the full problem, Eq.\ (\ref{Effective_Hamiltonian_final_1}) non-trivial.

The Hamiltonian Eq.\ \eqref{Effective_Hamiltonian_i_0} centered around a site $i \in B_{m}$ can be diagonalized by the following unitary transformation
\begin{equation}
\label{definition_f}
c_{j,\sigma}=\sum_{n=1}^{Z} \eta_{j n}^{(i)} f_{i,\sigma,n}\quad{\rm with}\quad j\in A_i \, .
\end{equation}
where $\hat \eta^{(i)}$ is a $Z\times Z$ unitary matrix and
\begin{equation}
\label{condition}
\eta_{jn}^{(i)} = \frac{1}{\sqrt{Z}} \quad \mbox{for $n=1$ and all $j$} \, .
\end{equation}
For a given $m$, unitarity of the $\hat \eta^{(i)}$ ensures that the annihilators $f_{i,\sigma,n}$ for all $i\in B_m$ and all $n$ obey the standard fermion anticommutation relations. 
With Eqs.\ \eqref{definition_f} and \eqref{condition} the effective Hamiltonian Eq.\ \eqref{Effective_Hamiltonian_i_0} adopts the following form:
\begin{eqnarray}
\label{Effective_Hamiltonian_1}
{\cal H}_{\rm eff}^{(i)}=-\frac{Z^2}{4}\sum_\sigma   f_{i,\sigma,1}^\dagger f_{i,-\sigma,1}f_{i,-\sigma,1}^\dagger f_{i,\sigma,1} \, .
\end{eqnarray}
It is expressed in terms of annihilators and creators referring to the ``bonding'' orbital $n=1$ only, which is the orbital that is symmetrically centered around the site $i \in B_{m}$.
 
We define the spin of the symmetric orbital
\begin{equation}
\label{definition_S}
{\bm{\mathcal S}}_{i,1}=\frac{1}{2}\left(f_{i,\uparrow,1}^\dagger,f_{i,\downarrow,1}^\dagger\right) \cdot {\boldsymbol \sigma} \cdot \left(f_{i,\uparrow,1},f_{i,\downarrow,1}\right)^T \, .
\end{equation}
For a one-dimensional system, ${\bm{\mathcal S}}_{i,1}$ is the spin of the bonding orbital made up by the two basis orbitals of the $A$-sublattice sites neighboring the $B$-sublattice site $i$. 
For two dimensions, it is a plaquette spin operating on the $A$-sublattice sites neighboring the $i$-th Kondo singlet (see Fig.\ \ref{fig:plaquette}). 
Using the following expression for the bond-spin, plaquette-spin, etc.\ operators 
\begin{equation}
{\bm{\mathcal S}}_{i,1}^2=\frac{3}{4} \sum_\sigma f_{i,\sigma,1}^\dagger f_{i,-\sigma,1}f_{i,-\sigma,1}^\dagger f_{i,\sigma,1} \, ,
\end{equation}
we can write the effective Hamiltonian in the conceptually very simple form
\begin{equation}
\label{Effective_Hamiltonian_final} 
{\cal H}_{\rm eff} 
=
- \frac{Z^2}{3} \alpha \sum_{m=1}^Z \sum_{i \in B_m} {\bm{\mathcal S}}_{i,1}^2
\, .
\end{equation}

For a given $m$, the ground state of the Hamiltonian Eq.\ \eqref{Effective_Hamiltonian_m_0} is a tensor product of the ground states of the $\tilde{\cal H}_{\rm eff}^{(i)}$ with $i\in B_{m}$. Since ${{\bm{\mathcal S}}_{i,1}^2=\frac{3}{4} (f_{i,\uparrow,1}^\dagger f_{i,\uparrow,1} - f_{i,\downarrow,1}^\dagger f_{i,\downarrow,1})^2}$, the eigenvalues of $\tilde{\cal H}_{\rm eff}^{(i)}$ are $-Z^2/4$ and $0$, and a ground state of $\tilde{\cal H}_{\rm eff}^{(i)}$ is characterized by a fully developed magnetic moment on a bond, plaquette etc.
Therefore, a state with all non-local spins aligned in, say, the $+z$ axis (spin up), not only constitutes a ground state of ${\cal H}_{\rm eff}^{(m)}$ for a particular $m$ but is obviously also a ground state of ${\cal H}_{\rm eff}$.
This proves that, at half filling and in the strong-coupling limit, the fully polarized state is among the ground states of the depleted Kondo or Anderson lattice.

\section{Further analysis of the effective Hamiltonian}
\label{sec:analysis}

To address the question of a possible ground-state degeneracy and fillings off half-filling, one may apply variational techniques and exact diagonalization. 
To this end it is convenient to assume periodic boundary conditions and to rewrite the Hamiltonian Eq.\ \eqref{Heff_0} in momentum representation:
\begin{eqnarray}
\label{Heff_k}
{\cal H}_{\rm eff}&=&\sum_{{\bf k},\sigma} E({\bf k}) c_{{\bf k},\sigma}^\dagger c_{{\bf k},\sigma}  
\nonumber \\
&+&
\frac{1}{L_A}\sum_{{\bf p}, {\bf q},{\bf k}}U_{{\bf p},{\bf q},{\bf k}}
c_{{\bf p},\uparrow}^\dagger c_{{\bf p}-{\bf k},\uparrow} c_{{\bf q},\downarrow}^\dagger c_{{\bf k}+{\bf q},\downarrow}  \, ,
\end{eqnarray}
where $L_A$ is number of $A$-lattice sites and where
\begin{equation}
E({\bf k}) = -\frac{D\alpha }{2}\omega^2({\bf k})
\label{eq:effenergies}
\end{equation}
is the effective dispersion.
Furthermore, 
\begin{equation}
  U_{{\bf p},{\bf q},{\bf k}} = \frac{\alpha}{2}\omega({\bf p})\omega({\bf q})\omega({\bf p}-{\bf k})\omega({\bf k}+{\bf q}) 
\end{equation}
are the parameters of the effective interaction among the electrons on the $A$-sublattice sites. 
If the original $D$-dimensional lattice is hypercubic ($D=1$ chain, $D=2$ square lattice and $D=3$ cubic lattice), the parameters can be expressed in terms of 
\begin{eqnarray}
  \omega(k) &=& 2 \cos\left(k/2 \right) \quad (D=1)
  \nonumber \\
  \omega(\ff k) &=& 
  4 \cos\left(k_{x}/2 \right)
  \cos\left(k_{y}/2 \right)
  \quad (D=2)
  \nonumber \\
  \omega(\ff k) &=& 
  4 \cos\left(k_{x}/2 \right)
  \cos\left(k_{y}/2 \right)
  + 2\cos(k_{z})
  \quad (D=3)
  \nonumber \\
\end{eqnarray}
Recall that the effective Hamiltonian ${\cal H}_{\rm eff}$ operates on the $A$-sublattice sites only. 
Therefore, the summations in Eq.\ \eqref{Heff_k} extend over the $\bf k$-points of the Brillouin zone corresponding to the $A$ sublattice which, e.g., is a square lattice (with different lattice constant) for $D=2$ but a b.c.c.\ lattice for $D=3$. 

One can easily check that the total particle-number ${\cal N}_A$, the total spin ${\mathbfcal S}_A$ and the total momentum operator ${{\mathbfcal Q}_A=\sum_{{\bf q},\sigma} {\bf q} \,c_{{\bf q},\sigma}^\dagger c_{{\bf q},\sigma}}$ are mutually commuting and commuting with ${\cal H}_{\rm eff}$. Correspondingly, the total particle number $N_A$, the total magnetization $M_A$ and the total momentum ${\bf Q}_A$ are conserved quantum numbers.

\subsection{Single spin-flip}
 
We first test the stability of the fully polarized state
\begin{equation}
\label{FP_state}
|\Psi_{\rm FP}\rangle =\bigotimes_{\bf k}^{E({\bf k})\leq\mu} c_{{\bf k},\uparrow}^\dagger |0 \rangle
\end{equation}
against a single spin flip for arbitrary filling, i.e.\ for arbitrary $N_{A}$ with $0 \le N_{A} \le L_{A}$.
Stability against single spin flip is ensured if the energy of the state \eqref{FP_state} with $M_A = \frac{1}{2} N_A$ is lower or equal to the ground-state energy of ${\cal H}_{\rm eff}$ in the sector with the same particle number $N_A$ but with total magnetization $M_A = \frac{1}{2} N_A - 1$. 

We consider trial states spanned by the orthonormal basis states
\begin{equation}
|{\bf q},{\bf k} \rangle =c_{{\bf k}+{\bf q},\downarrow}^\dagger c_{{\bf k},\uparrow}|\Psi_{\rm FP}\rangle  \, 
\label{eq:var}
\end{equation}
with arbitrary $\ff q$ and with $\ff k$ referring to occupied states, i.e.\ $E({\bf k})\leq\mu$.
The dimension of the corresponding Hilbert space sector is $L_A N_A$.
Note that at half-filling, i.e.\ $N_{A} = L_{A}$, all $\uparrow$-states are occupied.
Furthermore, it is worth mentioning that trivial degeneracies (apart from the spin degeneracy resulting from $SU(2)$ symmetry) arise off half-filling, namely if the total particle number $N_A$ is such that the highest occupied energy levels are not completely occupied.
For $D=1$, this is the case for even $N_{A}$.
Here, $|\Psi_{\rm FP}\rangle$ is not unique.
For those cases we have checked that it is sufficient to test the stability of one of the different fully polarized states.
 
To exploit total-momentum conservation, we make use of the block-diagonal structure of the Hamiltonian matrix ${\langle {\bf q}',{\bf k}'|{\cal H}_{\rm eff}|{\bf q},{\bf k}\rangle =\delta_{{\bf q}{\bf q}'}{\cal H}^{\bf q}_{{\bf k}'{\bf k}}}$. 
For a given $\ff q$, each block ${{\cal H}^{\bf q}_{{\bf k}'{\bf k}}}$ has the dimension $N_A$ and is diagonalized numerically by standard techniques (the largest system considered here has $N_A=10^3$).
We have performed calculations for dimension $D=1$ and fillings $0 \leq N_A \leq L_A$.

Our results can be summarized as follows:
For all fillings off half-filling, $0 < N_{A} < L_{A}$, there is only a single state with the same energy as the fully polarized state $|\Psi_{\rm FP}\rangle$ (in the sector with total magnetization $M_A=\frac{1}{2}N_A-1$) while all other states have higher energies. 
This is consistent with the expectation that, except of the trivial $2S_A+1$ spin degeneracy resulting from $SU(2)$ symmetry, $|\Psi_{\rm FP}\rangle$ is the unique ground state. (However, see next Sec.\ \ref{sec:ed}).

At half-filling the fully polarized state is a ground state.
For odd $N_{A}$, it is unique (apart from the trivial spin degeneracy). 
For even $N_A$ there are two states with the same energy as $|\Psi_{\rm FP}\rangle$ (in the sector with total magnetization $M_A=\frac{1}{2}N_A-1$). 
One trivially results from the SU(2) symmetry and has total spin $S_A = \frac{1}{2}L_A$. 
It is obtained as $\ca S_{A}^{-} |\Psi_{\rm FP} \rangle = (\ca S_{A}^x - i \ca S_{A}^{y}) |\Psi_{\rm FP} \rangle$.
There is another one, however, which has total spin $S_A = \frac{1}{2}L_A-1$.
Concluding, the fully polarized state is stable against a single spin flip, and there is a non-trivial ground-state degeneracy at half-filling only.

\subsection{Exact diagonalization}
\label{sec:ed}

To test these results, we have performed full exact-diagonalization studies and have calculated the exact ground state(s) of the effective Hamiltonian, Eq.\ \eqref{Heff_k}, for $D=1$ in the entire filling range.
We again make use of the block-diagonal structure of the effective Hamiltonian given by the conserved quantum numbers $N_A$, $M_A$ and $Q_{A}$. 
Models with up to ten $A$-sublattice sites can be treated easily in this way.

The results can be summarized as follows:
At half-filling the fully polarized state is a ground state.
For odd $N_{A}$ it is the unique ground state (apart from the trivial spin degeneracy).
For even $N_{A}=L_{A}$, there is one state with $S_A=\frac{1}{2}L_A-1$ which has the same energy as $|\Psi_{\rm FP}\rangle$. 
Including trivial degeneracies the ground-state degeneracy is thus given by $2L_A$. 

Off half-filling, for $N_{A} < L_{A}$ but still above quarter filling, $N_{A} > L_{A}/2$, the fully polarized state is the unique ground state for odd $N_{A}$, apart from spin degeneracy.
In this case the total ground-state momentum $Q_{A}=0$. 
For even $N_{A}$, it is still a ground state but there is an additional trivial two-fold degeneracy as there are two orthogonal ground states with momenta $Q_{A} = \pm \pi N_A/L_A$.
Therewith, the exact-diagonalization studies fully support the physical picture obtained from the variational approach. 

This is different, however, for fillings below quarter filling, i.e.\ for $N_{A} \le L_{A}/2$:
Still the fully polarized state is the unique ground state (apart from the trivial spin degeneracy) if $N_{A}$ is odd.
For even $N_{A}$, however, the unique ground state is a spin singlet, $S_{A}=0$.
This means that as a function of the total particle number $N_{A}$, the total spin oscillates between $S_{A} = N_{A}/2$ and $S_{A}=0$. 
Such behavior cannot be captured by the stability analysis described in the preceding section. 
This physics is rather unexpected.
A systematic and detailed analysis of the properties of the total spin-singlet states and the reason for the oscillations in $S_{A}$ will be addressed in a future publication.

\section{Diluted systems}
\label{sec:dilute}

\begin{figure}[bpt]
\centerline{\includegraphics[width=0.45\textwidth]{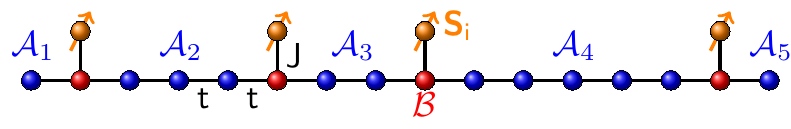}  }
\centerline{\includegraphics[width=0.45\textwidth]{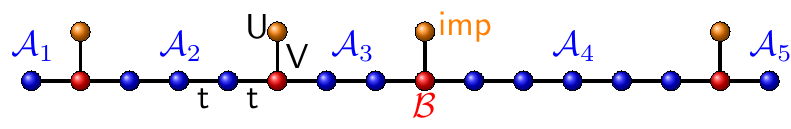}  }
\caption{(Color online) Schematic structure of a diluted Kondo lattice (top) and of a diluted Anderson lattice (bottom) for $D=1$.
The system of conduction-electrons sites consists of sites $\ca B$ (red) coupled, via $J$ or $V$, respectively, to the impurity sites, and of the remaining sites $\ca A$ (blue).
The set of $\ca B$ sites divides the $\ca A$ sites in different sets $\ca A_{\gamma}$.
In the strong-coupling limit $J\to \infty$ or $V \to \infty$, respectively, each conduction electron is confined to a certain region $\ca A_{\gamma}$.
The sets $A_{\gamma}$ may contain different numbers of $\ca A$ sites for each $\gamma$.
} 
\label{fig:depleted_ext_1D}
\end{figure}

The discussion has been done for a distance $d=2$ between the impurities so far but can straightforwardly be generalized to $d >  2$ and even to arbitrary impurity configurations, e.g., diluted systems with very few impurities and systems with reduced or absent translational symmetries. 
We continue the discussion for arbitrary lattice dimension $D$. 
Typical examples for one-dimensional systems are sketched in Fig.\ \ref{fig:depleted_ext_1D}.

For $d>2$, a conduction electron is no longer localized at a {\em single} $A$ site only but its motion is confined to a certain region $\ca A_{\gamma}$ in the strong-coupling limit.
Accordingly, the set of conduction-electron sites is divided in different groups: 
Sites coupled via $J$ (in the case of Kondo impurities) or via $V$ (Anderson impurities) belong to the group of sites $\cal B$.
The remaining sites belong to $\ca A$. 
Furthermore, $\ca A$ is partitioned into sets $\ca A_{\gamma}$ where, for each $\gamma$ the sites belonging to $\ca A_{\gamma}$ are coupled via the hopping term of the Hamiltonian (see Fig.\ \ref{fig:depleted_ext_1D}).
The considerations also comprise the single-impurity case ($R=1$) as a limit.
In this case and for $D>1$, no further partitioning of $\ca A$ is necessary.

There are three different energy scales to be considered. 
The largest energy scale is the excitation energy of one of the local singlets that are formed by the $\ca B$ and the impurity sites. 
This energy is of the order of $J$ (for the Kondo impurities) or $V$ (Anderson impurities). 
The second-largest energy scale is given by the hopping amplitude $t$ of the conduction electrons and is associated with the delocalization of the conduction electrons in each region ${\cal A}_\gamma$. 
The smallest energy scale corresponds to the motion of conduction electrons through the local singlets which is accompanied by {\em virtual} excitations of the singlets. 

This energy scale can be determined by degenerate perturbation theory.
To this end we decompose the Hamiltonian ${\cal H}$ in the following way:
\begin{eqnarray}
\label{HH0HtH1}
{\cal H}={\cal H}_0 + {\cal H}_t + {\cal H}_1 \, .
\end{eqnarray}
Here ${\cal H}_0$ describes the local singlets and is given by Eq.\ \eqref{H0_Kondo} for the case of Kondo impurities and by Eq.\ \eqref{H0_Anderson} for Anderson impurities. 
The second term, 
\begin{equation}
\label{Ht}
{\cal H}_t =t \sum_{\langle i, j \rangle \in {\cal A} } \sum_\sigma c_{i,\sigma}^\dagger c_{j,\sigma} 
\: ,
\end{equation}
is the nearest-neighbor hopping of the conduction electrons {\em within} the different sets ${\cal A}_\gamma$. 
The problems associated with ${\cal H}_0$ and with ${\cal H}_t$ are easily solved separately, and furthermore we have $[{\cal H}_0, {\cal H}_t]=0$. We will thus consider ${{\cal H}_0+{\cal H}_t}$ as the unperturbed Hamiltonian while 
\begin{equation}
\label{H1_extended}
{\cal H}_1 =t \sum_{i \in {\cal B}} \sum_{j \in \ca A}^{{\rm n.n. of \:} i} \sum_\sigma \left(c_{i,\sigma}^\dagger c_{j,\sigma} + \mbox{H.c.} \right) 
\end{equation}
is treated as the perturbation. 

Similar to Eq.\ \eqref{GS_H0}, the degenerate ground states of the unperturbed Hamiltonian ${{\cal H}_0+{\cal H}_t}$ can be written as:
\begin{equation}
|\phi_0\rangle=\bigotimes_{i \in \ca B} |{\rm LS}_i\rangle \bigotimes_{\gamma} |X_\gamma \rangle \, .
\end{equation}
Here $|{\rm LS}_i \rangle$ denotes a local Kondo singlet $|{\rm KS}_i\rangle$ or Anderson singlet $|i; 2,0,0 \rangle$, respectively, and $|X_\gamma \rangle$ denotes the Fermi sea of the system of conduction electrons on the sites $\cal A_\gamma$.
The filling of each of these Fermi seas must be determined by minimization of the total energy.

Apart from extreme cases with one or more completely filled or empty regions $\cal A_\gamma$, one has to perform fourth-order perturbation theory to lift the macroscopic degeneracy of the ground-state energy.
Contrary to the perturbation theory for the case $d=2$, and in addition to the virtual excitations of the local singlets, there are virtual excitations of the ``Fermi-sea'' ground states of the different subsystems $\cal A_\gamma$ as well.
These are described by the hopping term $\ca H_{t}$ and, therefore, their excitation energy is of the order of $t$. 
Formally, all calculations presented above must be repeated with a largely increased number of excitations differing in energy by $\ca O(t)$.
The corresponding excitation energies $E_{0} - E_{n}$ in the respective denominator of a term associated with a perturbative process, however, can be expanded in powers of $t$. 
At the order $t^{4}$ this does not lead to any correction and, apart from the hopping term itself, we therefore get the same result as before, i.e.:
\begin{equation}
\label{Heff_ext}
{\cal H}_{\rm eff}
=
{\cal H}_t - \frac{\alpha}{4} \sum_{i \in {\cal B}} \sum_{j_1,\ldots,j_4 \in \ca A_i}^{n.n.~{\rm of}~i}  \sum_\sigma c_{j_1,\sigma}^\dagger c_{j_2,-\sigma}c_{j_3,-\sigma}^\dagger c_{j_4,\sigma} \: .
\end{equation}
The expression for the effective coupling constant ($\alpha_K$ or $\alpha_A$) are also unchanged, see Eqs.\ (\ref{alpha_Kondo}) and (\ref{alpha_Anderson}).

For dimensions $D \ge 2$ the second term of the effective Hamiltonian \eqref{Heff_ext} describes interactions between sites belonging to different groups $\ca A_{\gamma}$ and $\ca A_{\gamma'}$, as for $D=1$. 
This applies to cases as shown in Fig.\ \ref{fig:plaquette}, for example (see also the $D=2$ system discussed in Ref.\ \onlinecite{STP13}). 
Nevertheless, for $D\ge 2$ and in the dilute limit, there is typically a {\em single} group $\ca A$ only. 
It is worth mentioning, however, that the effective interaction term in Eq.\ \eqref{Heff_ext} also connects different sites in the {\em same} group.
Furthermore, the term can be rewritten in the non-local spin representation again, and for $d>2$ the different non-local (bond, plaquette, etc.) spins commute with each other. 
One should note, however, that the problem is still non-trivial as the non-local spins do not commute with the hopping term ${\cal H}_t$.
The hopping term may in fact mediate e.g.\ magnetic correlations induced by the effective interactions over larger distances. 
A corresponding application showing cooperation of different magnetic exchange mechanisms has been discussed recently. \cite{SHP14}

\section{Discussion and conclusion}
\label{sec:con}

Using fourth-order degenerate perturbation theory we have derived an effective low-energy Hamiltonian $\ca H_{\rm eff}$ for the depleted Kondo lattice in the strong $J$ regime.
When $J\to \infty$, the main physical effect is the formation of local Kondo singlets at all sites where localized spins are coupled to the conduction-electron system.
These singlets are ``integrated out'', i.e., the localized spins and the corresponding sites of the conduction-electron system do not appear in the effective Hamiltonian as any excitation of a local Kondo singlet requires an energy of the order of $J \gg t$.
However, the effective Hamiltonian $\ca H_{\rm eff}$ still remembers their mere presence, and it causes the excess conduction electrons to scatter from 
the singlets.
This scattering effect is already included at the zeroth order in an expansion in powers of the hopping that links the sites where the local Kondo singlets are formed with the rest of the conduction-electron system. 

The first non-trivial correction is of fourth order and yields the effective interactions that are generated among the excess conduction electrons due to 
virtual excitations of the Kondo singlets.
Effective interactions therefore result from the internal structure of the local Kondo singlets and correlate the {\em a priori} non-interacting conduction-electron system. 
However, at fourth order, they are restricted to the nearest-neighbor sites of each of the singlets. 
For the Kondo impurity model in a semi-infinite chain geometry, Nozi\`eres \cite{Noz74,Noz76,NB80} already pointed out that a Hubbard-like interaction is induced.

We have explicitly carried out the fourth-order perturbation theory for depleted Kondo lattices with a spin-spin distance $d \ge 2$ (in units of the lattice constant). 
This also comprises the single-impurity case.
The effective interaction comes with a coupling constant $\alpha_K = 64 t^4 / 3 J^3$ and includes, besides the Hubbard term, a ferromagnetic Heisenberg exchange term, an antiferromagnetic isospin exchange, and a correlated hopping through the Kondo singlet.
The appearance of the ferromagnetic spin exchange is worth pointing out:
Its presence demonstrates that the Kondo effect not always competes with indirect magnetic coupling mechanisms that may promote ferromagnetism (such as RKKY) but, in the strong-$J$ limit, even generates ferromagnetic coupling which may induce ferromagnetic order eventually.
Finally, in dimensions $D>1$ additional three- and four-site interaction terms are obtained.

We found that the effective interaction at the site $i$ can be rewritten in a very compact and highly symmetric form as $- \alpha_{K} Z^{2} {\bm{\mathcal S}}_{i,1}^2/3$. 
Here, ${\bm{\mathcal S}}_{i,1}$ is the spin operator of the symmetric, bonding ($n=1$) orbital centered around the singlet at the site $i$ in the conduction-electron system ($Z$ is the coordination number).
Virtual excitations of the Kondo singlets thus favor the formation of a non-local conduction-electron spin moment in the nearest-neighbor shell around each singlet.

Usually, in local Fermi-liquid theory, \cite{Noz74,Col07} this term $\propto \alpha \propto t^{4} / J^{3}$ can safely be neglected against the scattering effect $\propto t$.
The effective interaction becomes important or even dominating, however, for depleted Kondo lattices where the non-local spins start to overlap. 
In the model with distance $d=2$ between the localized spins, there is in fact no hopping term at all in $\ca H_{\rm eff}$:
The excess conduction electrons are localized between the local Kondo singlets, and the effective interaction, via the Hubbard term, not only produces completely local spin moments in the conduction-electron system but also, via the Heisenberg term, couples them ferromagnetically. 
At the same time, the isospin and the correlated hopping terms are basically ineffective. 
This ``inverse indirect magnetic exchange'' has been seen to lead to a ferromagnetic ground state in DMRG calculations for the $D=1$ dimensional model at half-filling. \cite{STP13}

Here, we could prove analytically that the ground state is ferromagnetic (if non-degenerate). 
Namely, as indicated above, the $d=2$ depleted Kondo lattice reduces in the strong-$J$ limit to a spin-only lattice model of the form $\propto \sum_{i} {\bm{\mathcal S}}_{i,1}^2$. 
This model is still non-trivial as the non-local orbitals $|i, \sigma, 1\rangle = f^{\dagger}_{i,\sigma,1} | \mbox{vac.}\rangle$, to which the spins ${\bm{\mathcal S}}_{i,1}$ refer to, are just overlapping which makes the respective spins non-commuting.
However, by grouping the Kondo singlets and the associated non-local spins in sublattices such that overlap is avoided, the model on each individual sublattice is easily seen to have a ferromagnetic ground state. 
At half-filling, this rigorously proves that the fully polarized Fermi sea of electrons filled into the band deriving from the orbitals $|i, \sigma,1 \rangle$ is the ground state (or among the ground states in case of degeneracy).

Off half-filling, the magnetic properties of the one-dimensional depleted Kondo lattice with $d=2$ are not finally clarified. 
We could, however, get some insight by testing the fully polarized ferromagnetic state against a single spin flip as well as by exact-diagonalization (Lanczos) calculations for small systems with up to ten conduction-electron sites in the effective Hamiltonian.
The results can be summarized as follows: 
The fully polarized state is the unique ground state at half-filling and for fillings off half-filling but still above quarter filling (for odd total number of electrons; otherwise there is a small degeneracy). 
Below quarter filling, however, the (unique) ground state is ferromagnetic for an odd but a total spin singlet for an even number of excess conduction electrons.
This rather unexpected behavior awaits a physical explanation.
Further studies are under way, and results will be published elsewhere.

Finally, an only marginally more complicated perturbative analysis is necessary to treat the depleted Anderson lattice in the strong $V$ limit.
This limit is interesting as it produces the same effective low-energy model $\ca H_{\rm eff}$ at fourth order albeit with a different coupling constant $\alpha_{A} = t^4 (U^3+48UV^2) / 24V^6$.
As expected, this reduces to $\alpha_{K}$ in the (extended) Kondo limit \cite{SW66,SN02} where charge fluctuations are suppressed and the Schrieffer-Wolff transformation applies.
In other parameter regimes (but still for strong $V$) the coupling constant for the Anderson case is larger than that of the Kondo case, $\alpha_{A}>\alpha_{K}$, if compared at $J=8V^{2}/U$.
Comparing the results of DMRG calculations for both models ($D=1$, $d=2$, half-filling) in fact shows that the fully polarized ferromagnetic state, which is characteristic for the strong-coupling limit ($J$ or $V$, respectively), is approached earlier in the Kondo case.

\acknowledgments
We would like to thank Matthias Peschke for helpful discussions.
Support of this work by the Deutsche Forschungsgemeinschaft through the SFB 668 (project A14) is gratefully acknowledged.
All authors contributed equally to the paper.

\appendix 
\section{Eigenvectors and eigenvalues of the unperturbed Hamiltonian}
\label{App_Unperturbed_Hamiltonian}

Here, we present the eigenvalues and eigenvectors of the unperturbed Hamiltonian ${\cal H}_0^{(i)}$, which is a building block of the total unperturbed Hamiltonian. 
To label the orthogonal and normalized eigenvectors $|i;q,m,l\rangle$, we introduce the following quantum numbers: 
$q=0, \ldots 4$ is the total number of particles, and $m=0, \pm 1/2, \pm 1$ is the magnetic quantum number corresponding to the total spin.
Furthermore, $l$ enumerates states in the sector with given $q$ and $m$. 
One easily finds the following results:

(i) ground-state energy and ground state (spin and isospin singlet, non-degenerate):
\begin{eqnarray}
\nonumber
&&{\cal E}^{(i)}_{2,0,0}=-\frac{1}{4}\sqrt{U^2+64V^2} \\
\nonumber
&&|i;2,0,0\rangle= \frac{1}{\sqrt{2}}\cos\frac{\alpha}{2} \left(c_{i,\uparrow}^\dagger f_{i,\downarrow}^\dagger |i;0\rangle - c_{i,\downarrow}^\dagger f_{i,\uparrow}^\dagger |i;0\rangle \right) \nonumber \\
\nonumber
&& -\frac{1}{\sqrt{2}}\sin\frac{\alpha}{2} \left(f_{i,\uparrow}^\dagger f_{i,\downarrow}^\dagger |i;0\rangle +c_{i,\uparrow}^\dagger c_{i,\downarrow}^\dagger |i;0\rangle\right) 
\nonumber
\end{eqnarray}

(ii) first excited energy level (broken singlet states, 4-fold degenerate):
\begin{eqnarray}
\nonumber
&&{\cal E}^{(i)}_{1,\pm 1/2,0}={\cal E}^{(i)}_{3,\pm1/2,0}=-\frac{1}{4}\sqrt{U^2+16V^2} \\
\nonumber
&&|i;1,\sigma,0\rangle=\cos\frac{\beta}{2} f_{\sigma}^\dagger |i;0\rangle - \sin\frac{\beta}{2} c_{\sigma}^\dagger |i;0\rangle \\
\nonumber
&&|i;3,\sigma,0\rangle= \cos\frac{\beta}{2} c_{i,\uparrow}^\dagger c_{i,\downarrow}^\dagger f_{\sigma}^\dagger |i;0\rangle + \sin\frac{\beta}{2} c_{\sigma}^\dagger f_{i,\uparrow}^\dagger f_{i,\downarrow}^\dagger|i;0\rangle \nonumber \\
\nonumber
\end{eqnarray}

(iii) second excited level (spin triplet, 3-fold degenerate):
\begin{eqnarray}
\nonumber
&&{\cal E}^{(i)}_{2,0,1}={\cal E}^{(i)}_{2,\pm 1,0} =-\frac{1}{4}U \\
\nonumber
&&|i;2,0,1\rangle=\frac{1}{\sqrt{2}} \left(c_{i,\uparrow}^\dagger f_{i,\downarrow}^\dagger |i;0\rangle + c_{i,\downarrow}^\dagger f_{i,\uparrow}^\dagger |i;0\rangle \right)\\
\nonumber
&&|i;2,2\sigma,0\rangle=c_{\sigma}^\dagger f_{\sigma}^\dagger |i;0\rangle
\end{eqnarray}

(iv) third excited level (isospin triplet, 3-fold degenerate):
\begin{eqnarray}
\nonumber
&&{\cal E}^{(i)}_{2,0,2}={\cal E}^{(i)}_{0,0,0}={\cal E}^{(i)}_{4,0,0} =\frac{1}{4}U\\
\nonumber
&& |i;2,0,2\rangle=\frac{1}{\sqrt{2}} \left(c_{i,\uparrow}^\dagger c_{i,\downarrow}^\dagger |i;0\rangle - f_{i,\uparrow}^\dagger f_{i,\downarrow}^\dagger |i;0\rangle \right)\\
\nonumber
&& |i;0,0,0\rangle=|i;0\rangle \\
\nonumber
&& |i;4,0,0\rangle=c_{i,\uparrow}^\dagger c_{i,\downarrow}^\dagger f_{i,\uparrow}^\dagger f_{i,\downarrow}^\dagger |i;0\rangle
\nonumber
\end{eqnarray}

(v) fourth excited level (broken singlet states, 4-fold degenerate):
\begin{eqnarray}
\nonumber
&&{\cal E}^{(i)}_{1,\pm 1/2,1}={\cal E}^{(i)}_{3,\pm1/2,1}=\frac{1}{4}\sqrt{U^2+16V^2} \\
\nonumber
&&|i;1,\sigma,1\rangle=\sin\frac{\beta}{2} f_{\sigma}^\dagger |i;0\rangle + \cos\frac{\beta}{2} c_{\sigma}^\dagger |i;0\rangle \\
\nonumber
&&|i;3,\sigma,1\rangle= \sin\frac{\beta}{2} c_{i,\uparrow}^\dagger c_{i,\downarrow}^\dagger f_{\sigma}^\dagger |i;0\rangle - \cos\frac{\beta}{2} c_{\sigma}^\dagger f_{i,\uparrow}^\dagger f_{i,\downarrow}^\dagger|i;0\rangle \nonumber\\
\nonumber
\end{eqnarray} 

(vi) fifth excited level (spin and isospin singlet, non-degenerate):
\begin{eqnarray}
\nonumber
&&{\cal E}^{(i)}_{2,0,3}=-\frac{1}{4}\sqrt{U^2+64V^2} \\
\nonumber
&&|i;2,0,3\rangle= \frac{1}{\sqrt{2}}\sin\frac{\alpha}{2} \left(c_{i,\uparrow}^\dagger f_{i,\downarrow}^\dagger |i;0\rangle - c_{i,\downarrow}^\dagger f_{i,\uparrow}^\dagger |i;0\rangle \right) \nonumber \\
\nonumber
&& +\frac{1}{\sqrt{2}}\cos\frac{\alpha}{2} \left(f_{i,\uparrow}^\dagger f_{i,\downarrow}^\dagger |i;0\rangle +c_{i,\uparrow}^\dagger c_{i,\downarrow}^\dagger |i;0\rangle\right) 
\nonumber
\end{eqnarray}

In these expressions, $|i;0\rangle$ refers to the state where both, the conduction-electron site $i$ as well as the corresponding impurity site are empty. Furthermore, we have used the notation
\begin{eqnarray}
\nonumber
&& \cos\alpha \equiv \frac{U}{\sqrt{U^2+64V^2}}\, , \quad \sin\alpha = \frac{8V}{\sqrt{U^2+64V^2}} \, , \\ 
\nonumber
&& \cos\beta \equiv \frac{U}{\sqrt{U^2+16V^2}}\, , \quad \sin\beta = \frac{4V}{\sqrt{U^2+16V^2}} \, , 
\nonumber
\end{eqnarray}
and $\sigma=\uparrow,\downarrow$ corresponds to $\sigma=\pm 1/2$.

\end{document}